\documentclass{article}

\begin{document}

\begin{center}
  {\Large\bf Colour Reconnections and Rapidity Gaps}

  Leif L\"onnblad, NORDITA, Blegdamsvej 17, DK-2100 K{\o}benhavn {\O},
  Denmark, leif@nordita.dk
\end{center}

The models recently proposed by Buchm\"uller \cite{WB} and Ingelman
\cite{GI} to describe rapidity gap events in DIS at HERA are both
based on the assumption of additional colour exchange between the
struck system and the proton remnant in boson-gluon fusion events. In
this way the struck system may become colour singlet and a rapidity
gap may be formed between it and the remnant. In both models, the
colour exchange mechanism is assumed to be non-perturbative with
negligible momentum transfer. It is clear that such non-perturbative
interactions take place very late in the process, and cannot influence
the perturbative emissions in an event. Therefore the models rely
heavily on the assumption that the emission of hard gluons in the
forward region is suppressed, since any such emission would destroy a
potential gap.

\begin{figure}[b]
  \setlength{\unitlength}{0.063mm}
  \begin{picture}(1200,700)(0,-50)
    \put(50,100){\vector(1,0){1100}}
    \put(600,100){\vector(0,1){550}}
    \put(600,50){\makebox(0,0){(a)}}
    \put(100,100){\line(1,1){500}}
    \put(1100,100){\line(-1,1){500}}
    \put(410,310){\makebox(0,0){{\footnotesize $\bullet$}}}
    \put(410,310){\vector(1,-2){30}}
    \put(440,250){\vector(1,-1){60}}
    \put(500,190){\vector(3,-1){125}}
    \put(425,235){\makebox(0,0){{\footnotesize 1}}}
    \put(485,175){\makebox(0,0){{\footnotesize 2}}}
    \put(625,120){\makebox(0,0){{\footnotesize 3}}}
    \put(650,675){\makebox(0,0){{\footnotesize $\ln p_\perp^2$}}}
    \put(1175,70){\makebox(0,0){{\footnotesize $y$}}}
  \end{picture}
\setlength{\unitlength}{0.08bp}
\begin{picture}(2880,1728)(100,-100)
\put(2334,1414){\makebox(0,0)[r]{{\small\sc Lepto} 6.3}}
\put(2334,1514){\makebox(0,0)[r]{{\small\sc Ariadne} 4.07}}
\put(900,1464){\makebox(0,0)[r]{(b)}}
\put(1648,51){\makebox(0,0){$y$}}
\put(200,964){%
\makebox(0,0)[b]{\shortstack{$(1/N) dN/dy$}}%
}
\put(2697,151){\makebox(0,0){10}}
\put(2397,151){\makebox(0,0){8}}
\put(2098,151){\makebox(0,0){6}}
\put(1798,151){\makebox(0,0){4}}
\put(1499,151){\makebox(0,0){2}}
\put(1199,151){\makebox(0,0){0}}
\put(900,151){\makebox(0,0){-2}}
\put(600,151){\makebox(0,0){-4}}
\put(540,1519){\makebox(0,0)[r]{0.8}}
\put(540,1202){\makebox(0,0)[r]{0.6}}
\put(540,885){\makebox(0,0)[r]{0.4}}
\put(540,568){\makebox(0,0)[r]{0.2}}
\put(540,251){\makebox(0,0)[r]{0}}
\end{picture}

\vskip -5mm

  \caption[dummy]{{\it (a) The available phase space available for
  emission of a gluon with rapidity $y$ and transverse momentum
  $P_\perp$. (b) The rapidity distribution of gluons in the lab frame
  of HERA.}}

 \label{figphase}
\end{figure}
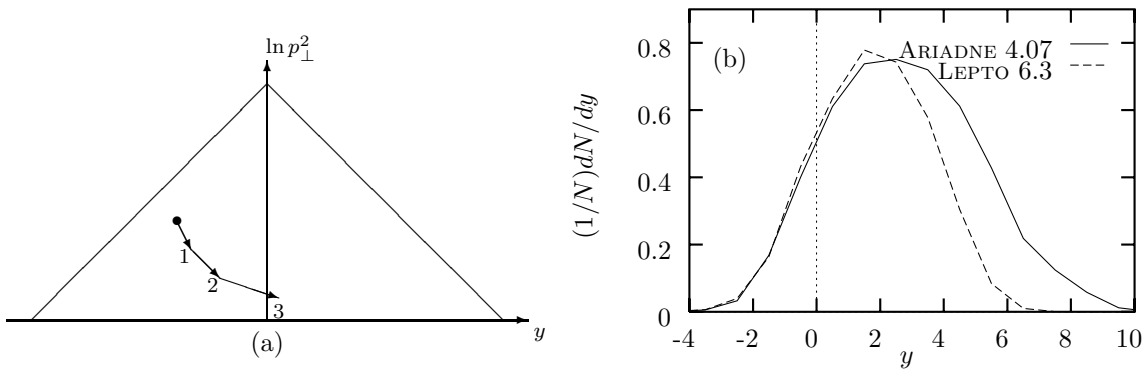

Naively one may think that additional gluon radiation is suppressed by
a power of $\alpha_S$, and indeed, the model of ref.~\cite{WB} does
not even consider it. However, the factor $\alpha_S$ is more than
compensated by the increase of phase space for such emissions at small
$x$, introducing large logarithms of $1/x$. In the model of
ref.~\cite{GI}, additional perturbative gluon emissions are taken into
account using a parton shower scenario based on the Altarelli-Parisi
or DGLAP \cite{DGLAP} evolution as implemented in the {\sc Lepto}
event generator program \cite{lepto}. With this model the rate and
distribution of rapidity gap events can be reproduced, provided the
parton shower is cut off at a virtuality scale around 4 GeV$^2$.

There are serious doubts whether at small $x$, the DGLAP evolution,
although being able to describe inclusive distributions like $F_2$, is
appropriate for the description of exclusive final state properties,
such as gluon emission. In the parton-shower language, the problem is
that the initial-state shower is strictly ordered in virtuality or
transverse momentum $p_\perp$. I.e.\ starting from the electro-weak
vertex, tracing successive emissions upwards in x towards the remnant
in rapidity, the $p_\perp$ decreases in each step as indicated in
fig.~\ref{figphase}a. At small $x$ and $Q^2$, where the bulk of events
are found at HERA, this means that the initial-state shower quickly
runs out of virtuality, preventing it from populating the forward
region in phase space. Consequently rapidity gaps occur frequently on
the partonic level, which may then survive the hadronization by
introducing the ``Soft Colour Interaction'' (SCI) of ref.~\cite{GI}.

There is no a priori reason why the emissions should be strongly
ordered. Indeed, in the BFKL \cite{bfkl} and CCFM \cite{CCFM}
evolution schemes, where large logarithms of $1/x$ are resummed,
which are believed to be more appropriate in the small-$x$ region, the
transverse momenta are unordered. Unfortunately, there is currently
no event generator implementing any of these schemes available. There
is, however, one generator -- {\sc Ariadne} \cite{ariadne} --
which implements an unordered cascade based on the Dipole Cascade Model
(DCM) \cite{dcm}. Recently, colour rearrangements of produced partons
was implemented in this program, introducing the possibility to get
large rapidity gaps in a similar way as in the models above, but with
an unordered cascade \cite{colrec}.

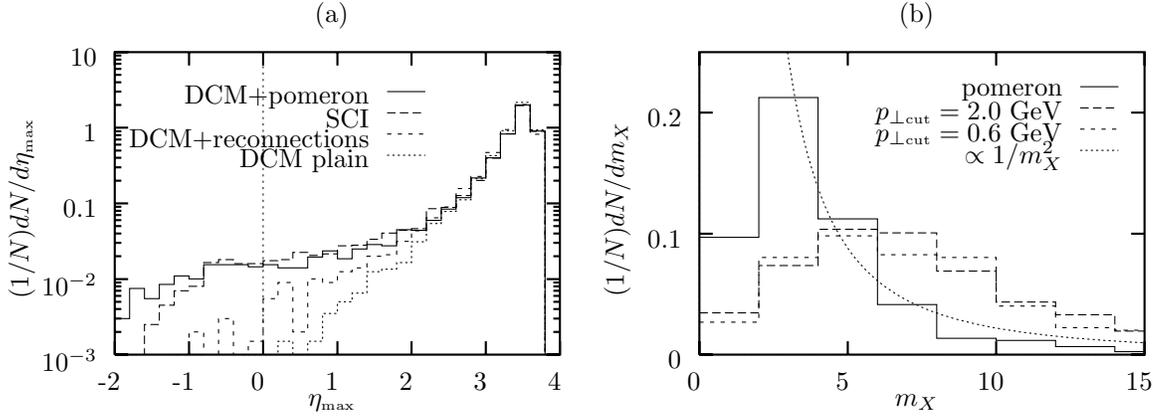
\begin{figure}
\setlength{\unitlength}{0.08bp}
\begin{picture}(2880,1728)(150,-600)
\put(1823,1191){\makebox(0,0)[r]{DCM plain}}
\put(1823,1291){\makebox(0,0)[r]{DCM+reconnections}}
\put(1823,1391){\makebox(0,0)[r]{SCI}}
\put(1823,1491){\makebox(0,0)[r]{DCM+pomeron}}
\put(1648,51){\makebox(0,0){$\eta_{\mbox{\tiny max}}$}}
\put(1648,1877){\makebox(0,0){(a)}}
\put(250,964){%
\makebox(0,0)[b]{\shortstack{$(1/N) dN/d\eta_{\mbox{\tiny max}}$}}%
}
\put(2697,151){\makebox(0,0){4}}
\put(2348,151){\makebox(0,0){3}}
\put(1998,151){\makebox(0,0){2}}
\put(1649,151){\makebox(0,0){1}}
\put(1299,151){\makebox(0,0){0}}
\put(950,151){\makebox(0,0){-1}}
\put(600,151){\makebox(0,0){-2}}
\put(540,1677){\makebox(0,0)[r]{$10$}}
\put(540,1321){\makebox(0,0)[r]{$1$}}
\put(540,964){\makebox(0,0)[r]{$0.1$}}
\put(540,607){\makebox(0,0)[r]{$10^{-2}$}}
\put(540,251){\makebox(0,0)[r]{$10^{-3}$}}
\end{picture}
\setlength{\unitlength}{0.08bp}
\begin{picture}(2880,1728)(400,-600)
\put(2334,1214){\makebox(0,0)[r]{$\propto 1/m_X^2$}}
\put(2334,1314){\makebox(0,0)[r]{$p_{\perp\mbox{\tiny cut}}=0.6$ GeV}}
\put(2334,1414){\makebox(0,0)[r]{$p_{\perp\mbox{\tiny cut}}=2.0$ GeV}}
\put(2334,1514){\makebox(0,0)[r]{pomeron}}
\put(1648,51){\makebox(0,0){$m_X$}}
\put(1648,1877){\makebox(0,0){(b)}}
\put(300,964){%
\makebox(0,0)[b]{\shortstack{$(1/N) dN/dm_X$}}%
}
\put(2697,151){\makebox(0,0){15}}
\put(1998,151){\makebox(0,0){10}}
\put(1299,151){\makebox(0,0){5}}
\put(600,151){\makebox(0,0){0}}
\put(540,1392){\makebox(0,0)[r]{0.2}}
\put(540,821){\makebox(0,0)[r]{0.1}}
\put(540,251){\makebox(0,0)[r]{0}}
\end{picture}
\vskip -2cm

  \caption[dummy]{{\it (a) The distribution of $\eta_{\mbox{\tiny
    max}}$, the largest pseudo-rapidity of any particle in the lab
    frame of a HERA detector. The full line is {\sc Ariadne} with the
    default pomeron inspired model, long-dashed is SCI, short-dashed
    is DCM with colour reconnections (CR) and the dotted line is plain
    DCM without CR or pomerons. (b) The distribution of $m_X$ for
    events at HERA with $\eta_{\mbox{\tiny max}}<3.2$. The full line
    is {\sc Ariadne} with pomerons, long-dashed DCM with CR and an
    increased cutoff, and short-dashed is the same with default
    cutoff.}}

  \label{figgap}
\end{figure}

The difference between the {\sc Ariadne} and {\sc Lepto} programs is
apparent in fig.~\ref{figphase}b, where the rapidity distribution of
gluons is shown for the two models. Clearly, {\sc Ariadne} gives much
more gluons in the forward region. This is reflected in the
$\eta_{\mbox{\small max}}$ distribution in fig.~\ref{figgap}a, where
the DCM with reconnections is far below.

But even if the number of gap events could be increased in {\sc
Ariadne} by simply increasing the cutoff in the cascade, thus reducing
the number of gluons, distributions such as the mass of the
diffractive system -- $m_X$, would not be well described by {\sc Ariadne}
because of the different distribution of gluons. This is clear from
fig.~\ref{figgap}b where the $m_X$ distribution is shown for two
different cutoffs and compared to the results from a pomeron-inspired
model \cite{aripom} which have the $1/m_X^2$ behavior seen in data.

In conclusion, the models \cite{WB,GI} which describe the rapidity-gap
events found in DIS at HERA, do so because they inhibit gluon
radiation in the forward region. There is no physical motivation why
this should be the done. Using an unordered cascade, where the effect
of the increase of available phase space with $\log 1/x$ is taken into
account, give qualitatively different distributions in possible
rapidity-gap events and can only explain data using a pomeron inspired
model, where the forward radiation is inhibited because of the initial
colour structure of the target proton. The models \cite{WB,GI} are
also based on the assumption of having colour exchanged without
exchanging momentum. Note that this exchange means that a charge going
very forward in rapidity is suddenly ``infinitely decelerated'' to go
backward in rapidity. It is questionable whether such an acceleration
of a charge is allowed without having accompanying radiation, which
would both require a momentum transfer, and most likely fill up the
potential rapidity gap.


\begin{thebibliography}{99}
\itemsep -1.0mm

\bibitem{WB}
{\sc {W.\ Buchm\"uller, A.\ Hebecker}},
\newblock {\em Phys.\ Lett.} {\bf B355} (1995) 573.
\bibitem{GI}
{\sc {A.\ Edin, G.\ Ingelman, J.\ Rathsman}},
\newblock {DESY 95--145}, July 1995.

\bibitem{DGLAP}
{\sc {V.N.\ Gribov, L.N.\ Lipatov}},
\newblock {\em Sov.\ J.\ Phys.} {\bf 15} (1972) 438 and 675;\\
{\sc {L.N.\ Lipatov}},
\newblock {\em Sov.\ J.\ Phys.} {\bf 20} (1975) 94;\\
{\sc {G.\ Altarelli, G.\ Parisi}},
\newblock {\em Nucl.\ Phys.} {\bf B126} (1977) 298;\\
{\sc {Yu.L.\ Dokshitser}},
\newblock {\em Sov.\ Phys.\ JETP} {\bf 46} (1977) 641.

\bibitem{lepto}
{\sc {G.\ Ingelman}},
\newblock {LEPTO version 6.3},
\newblock in {\em {Physics at HERA}}, eds,\ {\sc {W.\ Buchm{\"u}ller, G.\
  Ingelman}}, vol.~3, p. 1366, DESY, 1991.

\bibitem{bfkl}
{\sc {V.S.\ Fadin, E.A.\ Kuraev, L.N.\ Lipatov}},
\newblock {\em Sov.\ Phys.\ JETP} {\bf 45} (1977) 199;\\
{\sc {Ya.Ya.\ Balitsky, L.N.\ Lipatov}},
\newblock {\em Sov.\ J.\ Nucl.\ Phys.} {\bf 28} (1978) 822.

\bibitem{CCFM}
{\sc {M.~Ciafaloni}},
\newblock {\em Nucl.\ Phys.} {\bf B296} (1988) 49;\\
{\sc {S.~Catani, F.~Fiorani, G.~Marchesini}},
\newblock {\em Phys.\ Lett.} {\bf B234} (1990) 339;
\newblock {\em Nucl.\ Phys.} {\bf B336} (1990) 18.

\bibitem{ariadne}
{\sc {L.\ L{\"o}nnblad}},
\newblock {{\sc Ariadne} version 4.07 program and manual},
\newblock {{\em Comput.\ Phys.\ Commun.} {\bf 71} (1992) 15}.

\bibitem{dcm}
{\sc {G.\ Gustafson}},
\newblock {\em Phys.\ Lett.} {\bf B175} (1986) 453;\\
{\sc {G.\ Gustafson, U.\ Pettersson}},
\newblock {\em Nucl.\ Phys.} {\bf B306} (1988) 746;\\
{\sc {B.\ Andersson et al.}},
\newblock {\em Z.\ Phys.} {\bf C43} (1989) 625.

\bibitem{colrec}
{\sc {L.\ L{\"o}nnblad}},
\newblock {CERN--TH/95--218 (to appear in {\em Z.\ Phys.\ C})}, August 1995.

\bibitem{aripom}
{\sc {L.\ L\"onnblad}},
\newblock {\em Z.\ Phys} {\bf C65} (1995) 285.

\end{thebibliography}
\end{document}